\documentclass[%
aps,prb,superscriptaddress,twocolumn,floatfix,10pt]{revtex4-2}
\usepackage[utf8]{inputenc}
\usepackage{amsmath,amssymb}
\usepackage{empheq}
\usepackage{float}
\usepackage{lipsum}
\usepackage{mathtools,cuted}
\usepackage{esvect}
\usepackage{multirow}
\usepackage{xcolor}
\usepackage{soul}
\usepackage[export]{adjustbox}
\usepackage{graphicx}% Include figure files
\usepackage{dcolumn}% Align table columns on decimal point
\usepackage{bm}% bold math
\usepackage{hyperref}% add hypertext capabilities
\usepackage{physics}
\usepackage[mathlines]{lineno}% Enable numbering of text and display math
\usepackage{tikz}
\usetikzlibrary{tikzmark}
\usepackage{booktabs}
\usepackage[capitalise]{cleveref}
\usepackage{bbm}
\usepackage[english]{babel}
\usepackage{orcidlink}

\begin{document}
	
	\title{Emergent $\mathbb{Z}$-type topology in a quasi-one-dimensional extended QWZ model}
	
    \author{Z.~F. Osseweijer}
    \affiliation{%
    Institute for Theoretical Physics, Utrecht University, 3584CC Utrecht, The Netherlands\\
    }%
    \author{L. Eek}
    \affiliation{%
    Institute for Theoretical Physics, Utrecht University, 3584CC Utrecht, The Netherlands\\
    }
    \author{C. Morais Smith}
    \affiliation{%
    Institute for Theoretical Physics, Utrecht University, 3584CC Utrecht, The Netherlands\\
    }%
 
	\date{\today}
	
	\begin{abstract}
    We investigate the emergence of zero-dimensional topological end states in nanoribbons described by the Qi–Wu–Zhang (QWZ) model and its extensions with longer-range couplings. While dimensional reduction from two to one dimension is often assumed to preserve the symmetry classification of the parent system, here an additional symmetry can emerge originating from the real-space geometry of the ribbon. This symmetry acts as a chiral symmetry, combining orbital and spatial transformations, and promotes the effective one-dimensional system from symmetry class D to class BDI. We demonstrate that the existence of such a symmetry depends both on the long and end termination of the ribbon and exhibits an even–odd effect with respect to ribbon width, revealing that the commonly studied rectangular ribbons constitute a special high-symmetry case. For the conventional QWZ model, we derive analytic expressions for the topological phase boundaries of finite-width nanoribbons and characterize the resulting hybridization-gap phases through ($\mathbb{Z}_2$) and winding-number invariants. We further show that extended QWZ models with longer-range couplings support phases with multiple topological end states and higher winding numbers. These phases arise through distinct mechanisms, including the hybridization of multiple edge modes inherited from higher-Chern-number bulk phases. Our results demonstrate that both long and end termination can fundamentally alter the topological classification of confined Chern insulators, highlighting the interplay between crystalline geometry, emergent symmetries, and dimensional reduction.
	\end{abstract}
	
	\maketitle{}
	
	%\tableofcontents
	
	\section{\label{sec:Intro}Introduction
	}
    Topological materials host robust boundary states that are protected by the symmetries of the system. The existence and classification of the topological phases in non-interacting systems depends crucially on both symmetry and dimensionality, as captured by the tenfold way \cite{ryu_topological_2010}. Symmetry and dimensionality are usually taken to be rigid and independent properties. Consequently, confining one of the directions in a $d$-dimensional system to obtain a quasi-$(d-1)$-dimensional one, is not expected to alter the symmetry characterization. However, this expectation sometimes fails because the effective symmetries of the system can be modified in this process \cite{linder_anomalous_2009, hasan_colloquium_2010}. In particular, boundary states at opposite boundaries are brought into proximity and can hybridize \cite{liu_oscillatory_2010, zhang_edge_2015}. As a result, the confined system not only inherits the topology of the higher-dimensional bulk, but can also exhibit emergent topological phases induced by this hybridization \cite{shan_effective_2010}.
	
	Dimensional reductions have been studied in both three- (3D) to two-dimensional (2D) transitions, such as in thin film Sb$_2$Te$_3$ \cite{hsieh_observation_2009}, Bi$_2$Te$_3$  \cite{hsieh_observation_2009}, Bi$_2$Se$_3$ \cite{shan_effective_2010, maisel_liceran_topology_2024}, and germanium \cite{singh_topological_2013} to name a few, and in 2D to one-dimensional (1D) transitions, such as observed in germanene \cite{eek_electric-field_2025, klaassen_realization_2025} and graphene nanoribbons \cite{rizzo_topological_2018, zhao_topological_2021}. In the latter case, a 2D material can be confined into a quasi-1D structure by cutting it into a nanoribbon.
	In these geometries, edge modes localized at opposite boundaries hybridize and open a \textit{hybridization gap} as the width decreases. Importantly, such gaps can themselves host topological phases, giving rise to localized end modes in the quasi-1D system \cite{potter_multichannel_2010, tewari_topological_2012, wakatsuki_majorana_2014, wang_topological_2018, cook_finite-size_2023, traverso_emerging_2024, balling-anso_transition_2026}.
	Recent work has shown that the manner in which a nanoribbon is cut and the resulting edge terminations can strongly influence the emergent symmetries and consequently the topology of the hybridization gap \cite{osseweijer_topology_2026}.
	
	Here, we will investigate these effects in the Qi-Wu-Zhang (QWZ) model. This model provides a minimal realization of a Chern insulator on a square lattice and can also be interpreted as a 2D topological superconductor \cite{qi_topological_2006}. Its simple two-band structure keeps the system analytically tractable, while introducing orbitals of different parity.
	
	Previous works have shown that nanoribbons of the QWZ model can host 1D topological phases \cite{potter_multichannel_2010, tewari_topological_2012, wakatsuki_majorana_2014, wang_topological_2018, cook_finite-size_2023, balling-anso_transition_2026}. Based on the tenfold way, after such a dimensional reduction the system would be expected to remain within symmetry class D, characterized by a $\mathbb{Z}_2$ topological invariant. However, a chiral symmetry can emerge, promoting the system to class BDI and allowing for a $\mathbb{Z}$ classification in terms of a winding number, as indicated in Table \ref{tab:dimred} \cite{ryu_topological_2010, chiu_classification_2016}. 
	This symmetry enhancement has been discussed in Refs.~\cite{tewari_topological_2012,cook_finite-size_2023,wang_topological_2018} and is similar to the emergent chiral symmetry observed in Haldane nanoribbons \cite{osseweijer_topology_2026}.
    
    In this work, we will investigate these effects in the Qi-Wu-Zhang (QWZ) model. This model provides a minimal realization of a Chern insulator on a square lattice and can also be interpreted as a 2D topological superconductor \cite{qi_topological_2006}. Its simple two-band structure keeps the system analytically tractable, while introducing orbitals of different parity.

    Previous work has shown that nanoribbons of the QWZ model can host 1D topological phases \cite{potter_multichannel_2010, tewari_topological_2012, wakatsuki_majorana_2014, wang_topological_2018, cook_finite-size_2023, balling-anso_transition_2026}. Based on the tenfold way, after such a dimensional reduction the system would be expected to remain within symmetry class D, characterized by a $\mathbb{Z}_2$ topological invariant. However, a chiral symmetry can emerge, promoting the system to class BDI and allowing for a $\mathbb{Z}$ classification in terms of a winding number, as indicated in Table \ref{tab:dimred} \cite{ryu_topological_2010, chiu_classification_2016}. 
    This symmetry enhancement has been discussed in Refs.~\cite{tewari_topological_2012,cook_finite-size_2023,wang_topological_2018} and is similar to the emergent chiral symmetry observed in Haldane nanoribbons \cite{osseweijer_topology_2026}.
    \begin{table}[h]
    \centering
    \begin{tabular*}{0.8\linewidth}{@{\extracolsep{\fill}} c c c c c c c @{}}
    \toprule
    Class & T & C & S & 1D & 2D & 3D \\
    \midrule
    D   & 0 & +1 & 0 & $\mathbb{Z}_2$ & \tikzmarknode{D2D}{$\mathbb{Z}$} & 0 \\
    BDI & +1 & +1 & 1 & \tikzmarknode{BDI1D}{$\mathbb{Z}_{\phantom{2}}$} & 0 & 0 \\
    \bottomrule
    \end{tabular*}
    \begin{tikzpicture}[overlay, remember picture]
    \draw[->, thick]
        ([xshift=-3pt]D2D.west) to ([xshift=3pt]BDI1D.east);
    \end{tikzpicture}
    \label{tab:dimred}
    \caption{A subsection of the tenfold way, with symmetry classes D and BDI, their symmetries and their topological classifications in dimensions $d=1,2,3$. The arrow indicates the symmetry enhancement when confining the QWZ model to 1D.}
    \end{table}

    In this work, we demonstrate that this emergent chiral symmetry depends on the geometry of the nanoribbon. In particular, its presence depends on the end terminations of the nanoribbon and it exhibits an even/odd effect with respect to the nanoribbon width. 
    This dependence has previously been overlooked, as prior studies commonly constructed ribbons by coupling 1D chains in rectangular geometries, which constitute a special case where chiral symmetry is present for all widths. Furthermore, we extend the QWZ model by including next-nearest-neighbor (NNN) or next-next-nearest-neighbor (NNNN) couplings and show how these extensions lead the emergence of phases with multiple topological end modes. In this context, we identify distinct mechanisms through which higher winding numbers can be realized.

    This paper is organized as follows. In Sec.~\ref{sec:model}, we introduce the model describing (extended) QWZ nanoribbons. In Sec.~\ref{sec:endstates}, we investigate the emergence of topological end modes for different nanoribbon terminations. In Sec.~\ref{sec:topo}, we characterize the nanoribbons in terms of symmetries and define the associated topological invariants. Furthermore, we present the corresponding phase diagrams. In Sec.~\ref{sec:extended}, we study the extended models and elaborate on the different mechanisms that result in higher winding numbers. Finally, we present our conclusions in Sec.~\ref{sec:Conclusions}.

    \begin{figure*}
        \centering
        \includegraphics{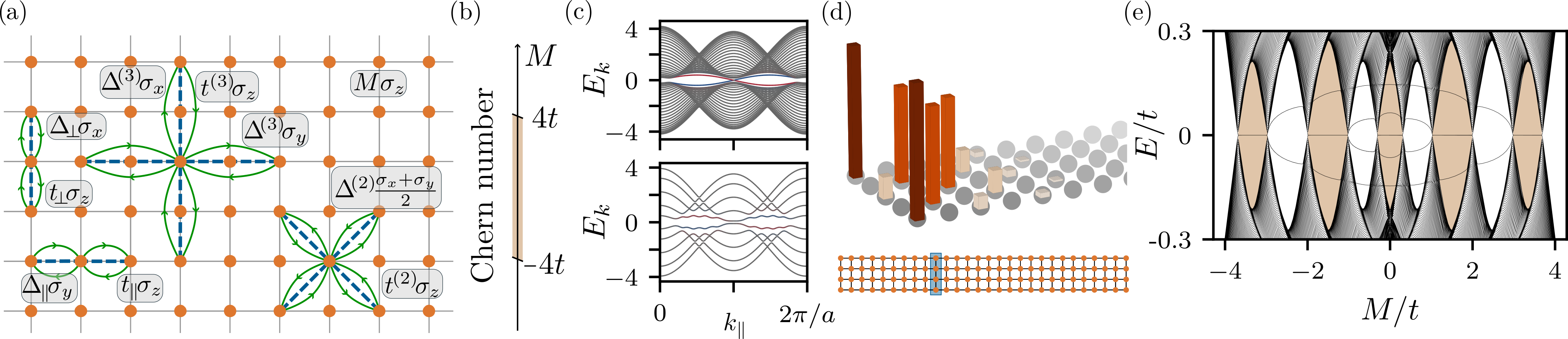}
        \caption{A 5-chain-wide parallel rectangular QWZ nanoribbon with $\Delta =0.22t$. (a) The 2D topological phase diagram. (b) The 2D lattice with the different hoppings $t$ (green) and couplings $\Delta$ (blue). (c) The spectrum of a 25- and 5-chain wide ribbon, with $t_\perp=t_\parallel=1$ and $\Delta_\perp=\Delta_\parallel=0.22t$ for a mass $M=0.2t$. (d) The wavefunction amplitude of a topological end mode at the left end of the ribbon for $M=0.03t$, together with the top view of a shorter ribbon. (e) The OBC spectra of a 250-site-long nanoribbon, as a function of $M$. The topological gaps are marked orange according to a non-trivial Zak phase.}
        \label{fig:5wideIntro}
    \end{figure*}

    \section{\label{sec:model}Extended QWZ models on nanoribbons}
    The QWZ model is a two-band tight-binding model on a 2D square lattice which hosts topological edge states. Depending on the context, the model describes a Chern insulator (CI) with an even and an odd parity orbital or it can been seen as a 2D generalization of the Kitaev chain \cite{qi_topological_2006, read_paired_2000}, hence a topological superconductor (TSC). In the former case, the spinors would be defined as $\Psi = (\Psi_+, \Psi_-)^T$, while in the latter case, they take the Nambu form $\Psi=(\Psi,\Psi^\dagger)^T$, and the Hamiltonian should be understood as a Bogoliubov-de Gennes (BdG) Hamiltonian acting in particle-hole space. In the remainder of this work, we will be aware of this duality. However, the results here hold independently of the interpretation. 
    Both models are captured by a 2D Hamiltonian, which is given by
    \begin{align}
        H_{Q} =& \left ( \sum_{\langle i\rangle} M c^\dag_{i}c_{i} + \sum_{\langle i,j\rangle_\parallel} t_{\parallel} c^\dag_{i}c_{j} + \sum_{\langle i,j\rangle_\perp}  t_{\perp}c^\dag_{i}c_{j} \right ) \sigma_z \nonumber \\
        +& \sum_{\langle i,j\rangle_\parallel}  \Delta_{\parallel} c^\dag_{i}c_{j} \sigma_y +  \sum_{\langle i,j\rangle_\perp} \Delta_{\perp} c^\dag_{i}c_{j} \sigma_x + \text{h.c.}
        \label{eq:hqwz}
    \end{align}
    Here, $(i,j)$ are site indexes and $\langle i,j\rangle_\parallel$ denotes the intra-chain nearest neighbors, and $\langle i,j\rangle_\perp$ the inter-chain nearest neighbors (NN). $\sigma_i$ is the $i$th Pauli matrix acting in the orbital basis (CI) or the particle-hole basis (TSC), $M$ is a mass term, $t_\parallel$ and $t_\perp$ are hopping terms parallel and perpendicular to the chains, and $\Delta_\parallel$ and $\Delta_\perp$ are spin-orbit coupling terms (CI) or Cooper pairing terms (TSC) within a chain and across chains, respectively. The extensions we will consider are
    $H = H_Q + H_{ext}$, with
    \begin{equation}
        H_{ext} = 2\sum_n c_n \left [\sum_{\langle i,j\rangle_n}t^{(n)} c^\dag_{i}c_{j} \sigma_z + \sum_{\langle i,j \rangle_n}\Delta^{(n)} c^\dag_{i}c_{j}\sigma_y \right ],
        \label{eq:hext}
    \end{equation}
    where the $\langle i,j\rangle_n$ is the set of $n$th order NNN, and $t^{(n)}$ and $\Delta^{(n)}$ capture the hopping and pairing between $n$th order neighbors. Note that there is no distinction in the direction anymore. Specifically, we will consider $n=2$, which includes first-order diagonal couplings and $n=3$, which includes second-order hoppings along the NN bonds, as illustrated in Fig.~\ref{fig:5wideIntro}(a). Here, the various couplings are indicated on a part of the bulk lattice. The $t$ hoppings terms and the $\Delta$ couplings are shown in green and blue, respectively.
    
    If $c_n = 0$ for all $n>1$, the original QWZ model is obtained and 2D topological phases occur when $\abs{M}<4|t|$ \cite{qi_topological_2006}. To investigate the topological phases of nanoribbons described by the QWZ model, we take open-boundary conditions in one direction. The Hamiltonian of such a system can be written as
    \begin{equation}
        h_N(k) = \begin{pmatrix}
            h_\parallel(k) & h_\perp & 0 & \dots & 0 \\
            h^\dag_\perp & h_\parallel(k) & h_\perp & \dots & 0 \\ 
            0 & h^\dag_\perp & h_\parallel(k) & \ddots & 0 \\
            \vdots & \vdots & \ddots & \ddots & h_\perp\\
            0 & 0 & 0 & h^\dag_\perp & h_\parallel(k)
            \label{eq:ribbonh}
        \end{pmatrix},
    \end{equation}
    where $h_\parallel$ contains the contributions in the periodic direction, 
    \[h_\parallel(k_x) = (M + 2t_\parallel \cos{k_x})\sigma_z + 2 \Delta_\parallel \sin k_x \sigma_y.\]
    $h_\parallel$ is now the Hamiltonian describing a Kitaev chain, and nanoribbons of the QWZ model can therefore be viewed as coupled Kitaev chains.
    These Kitaev chains are coupled to each other by
    \[h_\perp  = t_\perp\sigma_z + i\Delta_\perp \sigma_x. \]

    In Figs.~\ref{fig:5wideIntro}(b)-(c), the bulk and a wide ribbon described by the QWZ model are investigated. The 2D bulk phase diagram of the conventional QWZ model is given in Fig.~\ref{fig:5wideIntro}(a). Fig.~\ref{fig:5wideIntro}(c) shows the band spectra of a 25- and 5-chain wide ribbon, respectively, with two edge modes localized on the top (red) and bottom (blue) of the ribbons. In the 5-chain-wide case, the edge modes hybridize and a hybridization gap is opened. 
    
    \section{End modes \label{sec:endstates}}
    In the context of coupled Kitaev chains, this model has been extensively studied, and indeed it is known that it exhibits 1D topological phases characterized by robust 0D modes localized on the ends of the ribbons \cite{potter_multichannel_2010, tewari_topological_2012, wakatsuki_majorana_2014, wang_topological_2018, cook_finite-size_2023, balling-anso_transition_2026}. When the couplings in both directions are of the same strength, i.e. when $t_\perp=t_\parallel=1$, and $\Delta_\perp=\Delta_\parallel=\Delta$, we obtain the original QWZ model. In Figs.~\ref{fig:5wideIntro}(d)-(e), a 5-chain-wide and 250-site-long QWZ nanoribbon is investigated with $\Delta=0.22t$. In Fig.~\ref{fig:5wideIntro}(d), the wavefunction amplitude of a topological zero-energy mode at the end of a QWZ nanoribbon is shown. In Fig.~\ref{fig:5wideIntro}(e) the low-energy OBC spectra are shown for mass values within this 2D topological regime. Here, the opening of hybridization gaps is observed. These gaps close at specific values of $M$ and the system undergoes a topological phase transition. The gaps are alternatingly in the 1D trivial (white) and topological phase (orange), which have a Zak phase $\gamma = \pi$ and host the characteristic pinned zero-energy (Majorana) modes, in line with earlier results \cite{potter_multichannel_2010, tewari_topological_2012, wakatsuki_majorana_2014, wang_topological_2018, cook_finite-size_2023, balling-anso_transition_2026}. Furthermore, in-gap finite-energy \textit{satellite states} are observed. These non-topological states are also localized on the end of the ribbons, but their energy evolves with $M$, indicating their lack of symmetry protection. 
    \begin{figure*}
        \centering
        \includegraphics{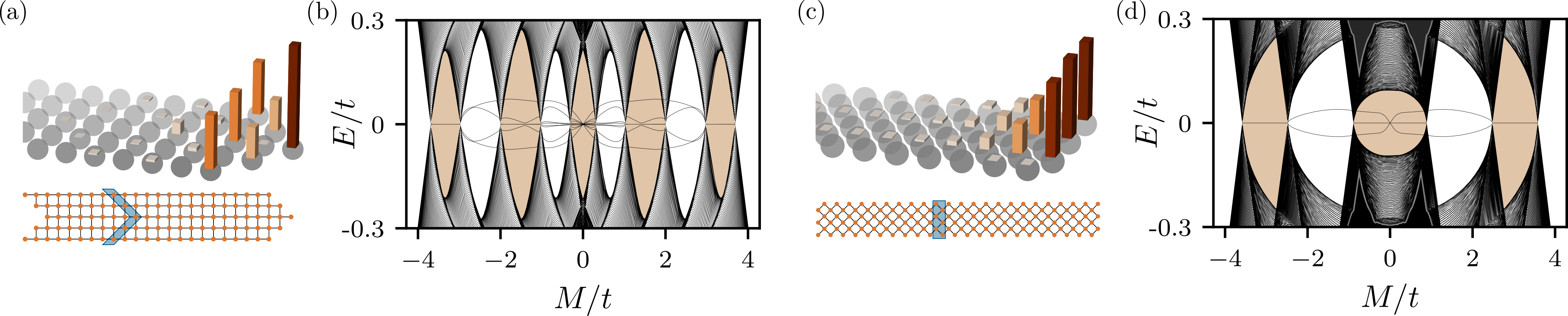}
        \caption{Two additional terminations of nanoribbons made from a 2D QWZ model. (a) and (b) A nanoribbon which long direction is parallel to the lattice and has a pointed sharp termination. (c) and (d) A nanoribbon which long direction is angled 45 degrees with respect to the lattice and has a rectangular termination. (a) and (c) The wavefunction amplitude of a topological mode at the ends of their respective nanoribbon, together with a top view of a shorter version of the lattice. (b) and (d) The OBC spectra of these nanoribbons as a function of $M$. Here, the topological hybridization gap are marked in orange according to a non-trivial Zak phase.}
        \label{fig:structures}
    \end{figure*}
    
    Up to this point, QWZ nanoribbons were constructed by coupling Kitaev chains, resulting in rectangular nanoribbons, with $t_\parallel$ parallel to its long direction. However, end termination can drastically alter the topological nature of similar hybridization gaps in honeycomb nanoribbons \cite{osseweijer_topology_2026}.
    Therefore, we switch perspectives from the coupling of chains to the cutting of quasi-1D structures from the 2D bulk, which allows for the formation of other structures. In Fig.~\ref{fig:structures}, two additional structures are shown. These terminations can also host topological modes, as illustrated in Fig.~\ref{fig:structures}. 
    Figs.~\ref{fig:structures}(a) and (b) depict a 5-chain-wide sharp cut nanoribbon with a sharp cut end termination in which the long direction is parallel to the lattice, and (c) and (d) depict a 6-chain-wide diagonal rectangular nanoribbon with a rectangular end termination, in which the long direction forms a 45 degree angle with respect to the lattice. 
    Both ribbons can host topological end modes as shown in Figs.~\ref{fig:structures}(a) and (c), respectively. Here, the wavefunction amplitude at the end of a long nanoribbon is plotted, together with the top view of a short nanoribbon to illustrate the termination of both ends. 
    In Figs.~\ref{fig:structures}(b) and (d), the low-energy OBC spectra are shown as a function of $M$. Again, multiple gap closing conditions are found with alternating topological and trivial character, as corroborated by the existence of pinned zero-energy modes. Once again, satellite states are found, the energy of which depends on the parameters. Compared to Fig.~\ref{fig:5wideIntro}(e), the energy dependence is different, further indicating that they are not symmetry protected.

    \section{\label{sec:topo}Topology}
	\subsection{Symmetries}
    Although the Zak phase is always defined for (quasi)-1D systems, it's value is not by definition quantized, in contrast with 2D topological invariants such as a Chern number. The Zak phase only quantizes under the constraint of a symmetry \cite{zak_berrys_1989}. Since the gaps of Figs.~\ref{fig:5wideIntro}(b), \ref{fig:structures}(b) and (d) yield a quantized Zak phase, there should be a symmetry that is responsible for this behavior. Indeed, the Bloch Hamiltonian of these nanoribbons hosts two distinct symmetries; particle-hole symmetry (PHS) \cite{qi_topological_2006} and an emergent chiral symmetry \cite{tewari_topological_2012, wakatsuki_majorana_2014}. 
    
    \paragraph{Particle hole symmetry} PHS is inherited from the original 2D bulk Hamiltonian. Consequently, it is independent of the geometry of the lattice. In the basis adopted here, it is given by the operator 
    \[ P = U_p\mathcal{K},\]
    where $U_p =(\mathbb{I}_n \otimes \sigma_x)$ is unitary and $\mathcal{K}$ represents complex conjugation, such that
    \[Ph(k)P^{-1} = -h(-k).\]
    Using PHS, we can quantify the topological phases using a Pfaffian method \cite{kitaev_unpaired_2001},
    \[ (-1)^\nu=\prod_{k=\{0,\pi\}} \text{sign}\left \{\text{Pf}[ h(k)U_p ] \right \}, \]
    where $\nu$ captures the $\mathbb{Z}_2$ topological nature, being either $0$ for a trivial phase or $1$ for a topological phase.
    \paragraph{Chiral symmetry} On top of PHS, these nanoribbons may have an emergent chiral symmetry, depending on the termination of the lattice. This symmetry acts both on the real-space and orbital degrees of freedom, to form a `flavored' inversion symmetry. For parallel rectangular ribbons, such as the one in Fig.~\ref{fig:5wideIntro}(d), it is present for all widths, but for the geometries of Fig.~\ref{fig:structures}, chiral symmetry only emerges in odd width nanoribbons. If present, this symmetry can be expressed in matrix form as
    \[ \Gamma = X \otimes i\sigma_x, \]
    where $X$ is the $n \times n$ matrix with ones on its anti-diagonal, and $\sigma_x$ is the Pauli $x$ matrix. This symmetry acts on the Bloch Hamiltonian as
    \[ \Gamma h(k) \Gamma^{-1} = -h(k), \]
    and therefore it behaves as a chiral symmetry. A similar even/odd effect in the occurrence of a flavored inversion symmetry acting as a chiral symmetry has been observed in honeycomb nanoribbons \cite{osseweijer_topology_2026}. 
    Combined with the aforementioned PHS, the existence of chiral symmetry implies a time-reversal like symmetry $\mathcal{T} = \mathcal{K}$ \cite{tewari_topological_2012, wakatsuki_majorana_2014}.  This places the odd-width (and even-width parallel rectangular) nanoribbons in class BDI, which has $\mathbb{Z}$-type topology, not captured by the $\mathbb{Z}_2$ Pfaffian invariant. In the presence of a chiral symmetry, one can calculate an integer valued winding number, which can quantify the $\mathbb{Z}$-type topology.

    \begin{figure*}
        \includegraphics[]{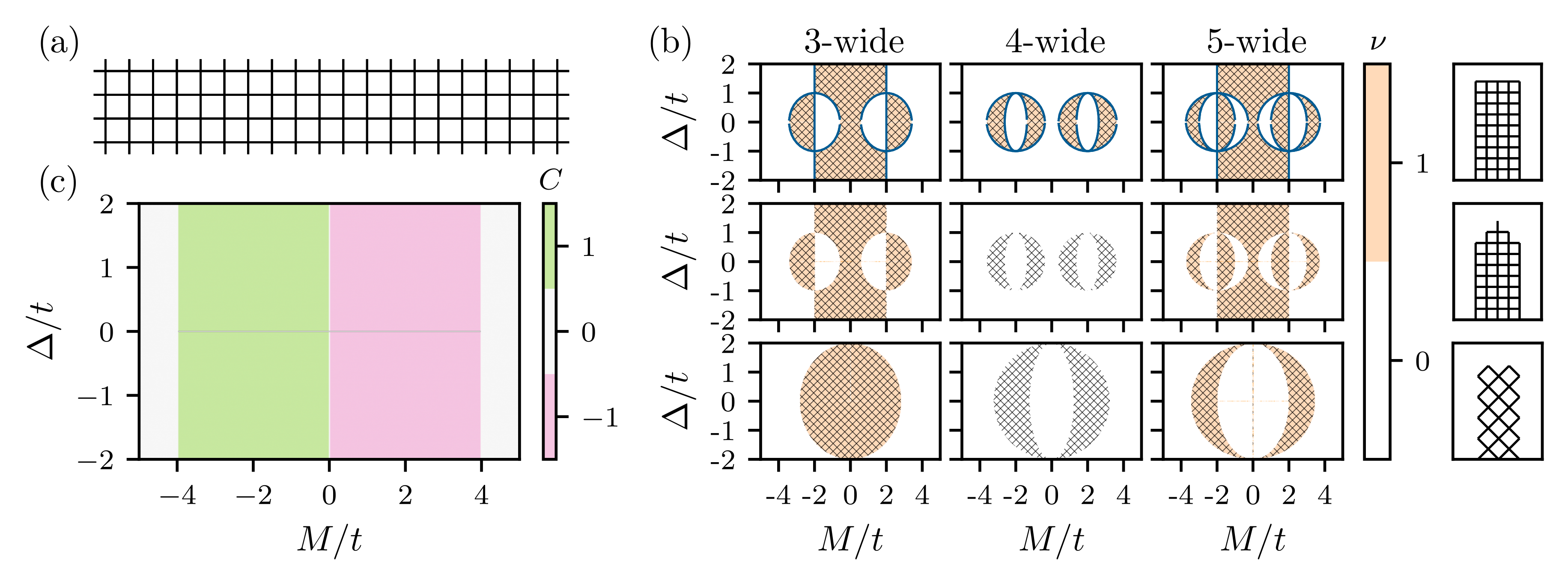}
        \caption{The topological phase space of (non-extended) QWZ nanoribbons ($c_{n>1} = 0, t_\parallel=t_\perp=t,\Delta_\parallel=\Delta_\perp=\Delta$). (a) The rectangular lattice. (b) The phase space of the QWZ nanoribbons, for 3-, 4-, and 5-site-wide parallel rectangular, parallel sharp and diagonal rectangular nanoribbons. Here, a non-zero Pfaffian invariant is indicated by hatched regions, and the winding number is captured by the colors. Note, 4-site-wide sharp and diagonal nanoribbons do not allow for a winding number, and consequently only host $\mathbb{Z}_2$ topology. (c) The 2D phase diagram showing a non-zero Chern number when $|M|<4t$. Furthermore, the phase boundaries derived in Eq.~\eqref{eq:PhaseBoundary} are shown for a rectangular nanoribbon. }
        \label{fig:phase}
    \end{figure*}
    
    To calculate the winding number, we transform the Bloch Hamiltonian using the eigenvectors $U$ of $\Gamma$ into its chiral basis
    \[\tilde{h}(k) = U^{-1} h(k) U = \begin{pmatrix}
        0 & q(k)\\
        q^\dag(k) & 0
    \end{pmatrix}.\]
    Now, the winding number is given in terms of $q(k)$ as
    \begin{equation}
        w = \frac{i}{2\pi}\oint_{BZ}dk  \log[\det(q)] \in \mathbb{Z} 
        \label{eq:winding}.
    \end{equation}
    This topological invariant can capture the full $\mathbb{Z}$-type topology, and is related to $\nu$ by 
    \[\nu = w \mod 2.\]
    However, this symmetry is not fully spectral, and requires a mirror axis in the long direction. This gives this symmetry a crystalline character and makes the protection it offers generally weaker.

    Finally, if $\Delta_\perp = 0$ the system becomes mirror symmetric along its short direction. This symmetry can be expressed as 
    \[ M = X \otimes I_2,\]
    where $I_2$ is the $2\times2$ identity on orbital basis (CI) or particle-hole basis (TSC). This symmetry can be used to enhance the chiral symmetry, allowing for further decomposition of $q^\dag(k)$ into a $q_+^\dag(k)$ and a $q_-^\dag(k)$ block. These blocks generally result in opposite winding, but do not cancel as they have different parity under this mirror symmetry. In these cases, the winding of $q_+(k)$ and $q_-(k)$ should be calculated independently and the total winding is
    \[ w = \abs{w_+} + \abs{w_-}.\]
    
    \subsection{Analytic phase boundaries}
    Because a topological invariant can only change when the gap closes, it is useful to calculate where the energy gap vanishes. For the (conventional) QWZ model of Eq.~\eqref{eq:hqwz}, we can derive these conditions analytically. The ribbon Hamiltonian in Eq.~\eqref{eq:ribbonh} is a block tridiagonal matrix. Although the internal spin structure prohibits direct diagonalization of the tridiagonal structure on the chain space, obfuscating the calculation of the eigenvalues, a recursive method can be applied to obtain the determinant. A gap closes when the bands around the Fermi level become degenerate, which due to PHS has to occur at $E=0$. Therefore, the gaps close when the determinant of $h_N(k)$ vanishes. Furthermore, due to the $\Delta_\parallel \sin(k_x)\sigma_y$ term, the gaps can only close at $k_x=\{0,\pi\}$. We define $m = M\pm2t_\parallel$, where the sign depends on the choice of $k_x$. With these considerations, Eq.~\eqref{eq:ribbonh} simplifies to
    \begin{equation}
        h_\parallel = m\sigma_z.
    \end{equation}
    The determinant of block tridiagonal matrices is given by \cite{molinari_determinants_2008}
    \[ \det(h_N) = D_N = \prod_{j=1}^N \det(\mathcal{F}_j),\]
    where $\mathcal{F}_j$ is a $2\times2$ matrix, determined by the recursion relation
    \begin{align*}
        \mathcal{F}_1 &= H_\parallel, \\
        \mathcal{F}_j &= H_\parallel - H^\dag_\perp \mathcal{F}^{-1}_{j-1} H_\perp.
    \end{align*}
    This is further simplified by realizing that $\mathcal{F}_j \propto \sigma_z$, such that the problem can be reduced to a scalar recursion, and the determinant is given by
    \begin{align*}
        D_N &= \prod_{j=1}^N -(\alpha_j)^2, & \alpha_j& = m - \frac{\det(F_1)}{\alpha_{j-1}} = m - \frac{(t_\perp^2-\Delta_\perp^2)}{\alpha_{j-1}},
    \end{align*}
    with $\alpha_1 = m$. This is a continued fraction, and to obtain a direct formula, we define
    \[ \alpha_j = \frac{p_j}{p_{j-1}},\] 
    such that $p_j= mp_{j-1} - (t_\perp^2-\Delta_\perp^2)p_{j-2}$, and $\det(h_N)=(-1)^N(p_N)^2$. Consistently with the boundary condition $a_1=M$, we choose $p_0=1$ and $p_1=M$. The next step is to assume $p_j \propto \lambda^j$, which yields
    \[ \lambda^2-m\lambda-(t_\perp^2 - \Delta_\perp^2) = 0, \]
    and which is solved by
    \begin{figure*}
        \includegraphics[]{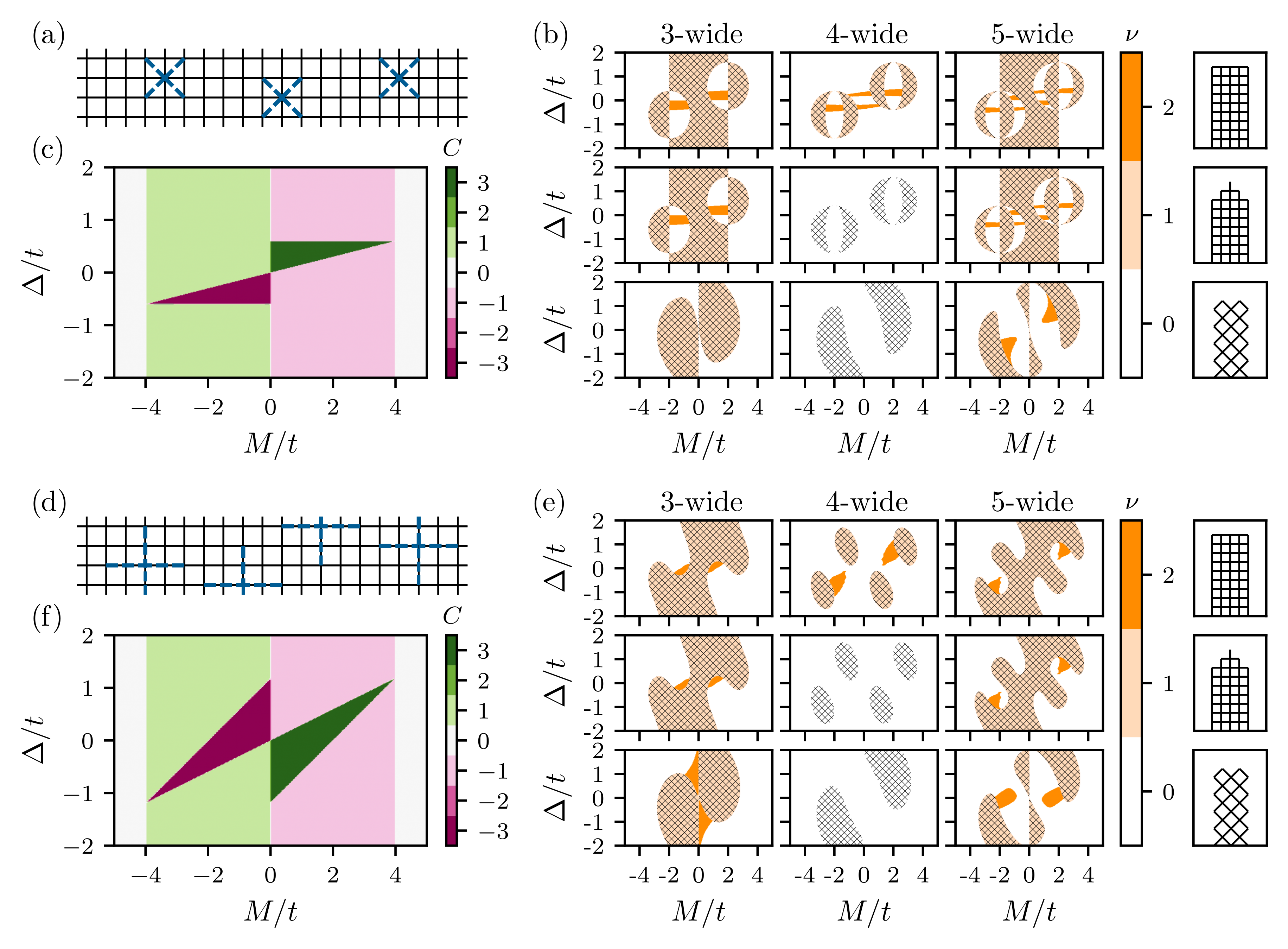}
        \caption{The topological phase space of extended QWZ nanoribbons ($t_\parallel=t_\perp=t,\Delta_\parallel=\Delta_\perp=\Delta$, $\Delta^{(n)}=0.2t$, $t^{(n)}=0$), with (a), (b) and (c) including NNN hopping $c_2 = 1$ and $c_{n\geq3}=0$ and (d), (e) and (f) including parallel NNNN hopping $c_3 =1$ and $c_{n\neq3}=0$. (a) The couplings considered when $c_2 = 1$. (b) The phase space of the nanoribbons, for 3-, 4-, and 5-site-wide parallel rectangular, parallel sharp and diagonal rectangular nanoribbons. Here, an non-zero Pfaffian invariant is indicated by hatched regions, and the winding number is captured by the colors. (c) The 2D phase diagram showing a non-zero Chern number when $|M|<4t$. (d), (e) and (f) Analogous to (a), (b) and (c) but with NNNN ($c_3 \neq 0 $) hopping instead of NNN hopping. }
        \label{fig:extendedHoppings}
    \end{figure*}
    \[ \lambda_\pm = \frac{m \pm \sqrt{m^2+4(t_\perp^2 - \Delta_\perp^2)}}{2}. \]
    A general expression of $p_j$ is now given by
    \[ p_j = c_+\lambda^j_+ + c_-\lambda^j_-, \]
    where $c_\pm$ can be obtained from the boundary conditions $p_0=1$ and $p_1=M$, such that
    \[ p_j = \frac{\lambda_+^{j+1}-\lambda_-^{j+1}}{\lambda_+-\lambda_-}. \]
    Finally, the gap closing condition is given by $p_N = 0$, and to solve $p_N = 0$, we define
    \begin{equation}
        m=2\sqrt{t_\perp^2 - \Delta_\perp^2}\cos{q} \quad q \in \mathbb{C}, 
        \label{eq:massCircle}
    \end{equation}
    such that $\lambda_\pm = \sqrt{(t_\perp^2-\Delta_\perp^2)}e^{\pm iq}$ and
    \[ p_j = \left ( \sqrt{(t_\perp^2-\Delta_\perp^2)} \right )^j \frac{\sin[(j+1)q]}{\sin(q)}. \]
    Solving $p_N = 0$ quantizes $q = n\pi/(N+1)$, $n = 1\dots N$, and for each value of $q$, a corresponding relation for the parameters is given by Eq.~\eqref{eq:massCircle}. After some algebraic manipulation, this becomes
    \begin{equation}
        \left ( \frac{M\pm2t_\parallel}{2\cos q} \right )^2+ \Delta_\perp^2 = t_\perp^2,
        \label{eq:PhaseBoundary}
    \end{equation}
    which describes ellipses in the $(M, \Delta_\perp)$-plane centered at $(M\pm2t_\parallel,0)$ with major and minor axes depending on $t_\perp$. When $q=\pi/2$, the cosine vanishes and the solution is instead given by $M=\pm2t_\parallel$, i.e. the gap closing conditions of the single chain. This only occurs for ribbons consisting of an odd number of chains, as the denominator $N+1$ has to be even. These phase boundaries are shown in Fig.~\ref{fig:phase}(b) and indeed they separate regions of different winding numbers.

    Finally, although direct diagonalization of the eigensystem is complicated, the $\cos q$ and $\sin q$ are closely related to the eigenvalues and -vectors of open tridiagonal matrices and $q$ can be interpreted as a discretized momentum.
    
    \subsection{Phase diagrams}
    Finally, we investigate the phase space and show the phases diagrams of the model, where we assume $t_\perp = t_\parallel=t$ and $\Delta_\perp = \Delta_\parallel=\Delta$, i.e. there is no distinction between the two directions. For now, conventional QWZ nanoribbons are considered, i.e.~$c=0$ in Eq.~\eqref{eq:hext}. Fig.~\ref{fig:phase} summarizes the topological phases in the bulk and on nanoribbons described by the conventional QWZ model. Fig.~\ref{fig:phase}(a), shows the square lattice formed by NN couplings. Fig.~\ref{fig:phase}(b) depicts the phase diagrams of 3-, 4-, and 5-chain-wide parallel rectangular, sharp, and diagonal QWZ nanoribbons, which highlights the importance of termination. Specifically, for 4-chain-wide, only the parallel rectangular ribbon of Fig.~\ref{fig:5wideIntro}(d), admits a winding number, although all terminations admit a Pfaffian invariant. Fig.~\ref{fig:phase}(c) shows the phase diagram of the 2D bulk in terms of a Chern number.

    \begin{figure*}
        \centering
        \includegraphics{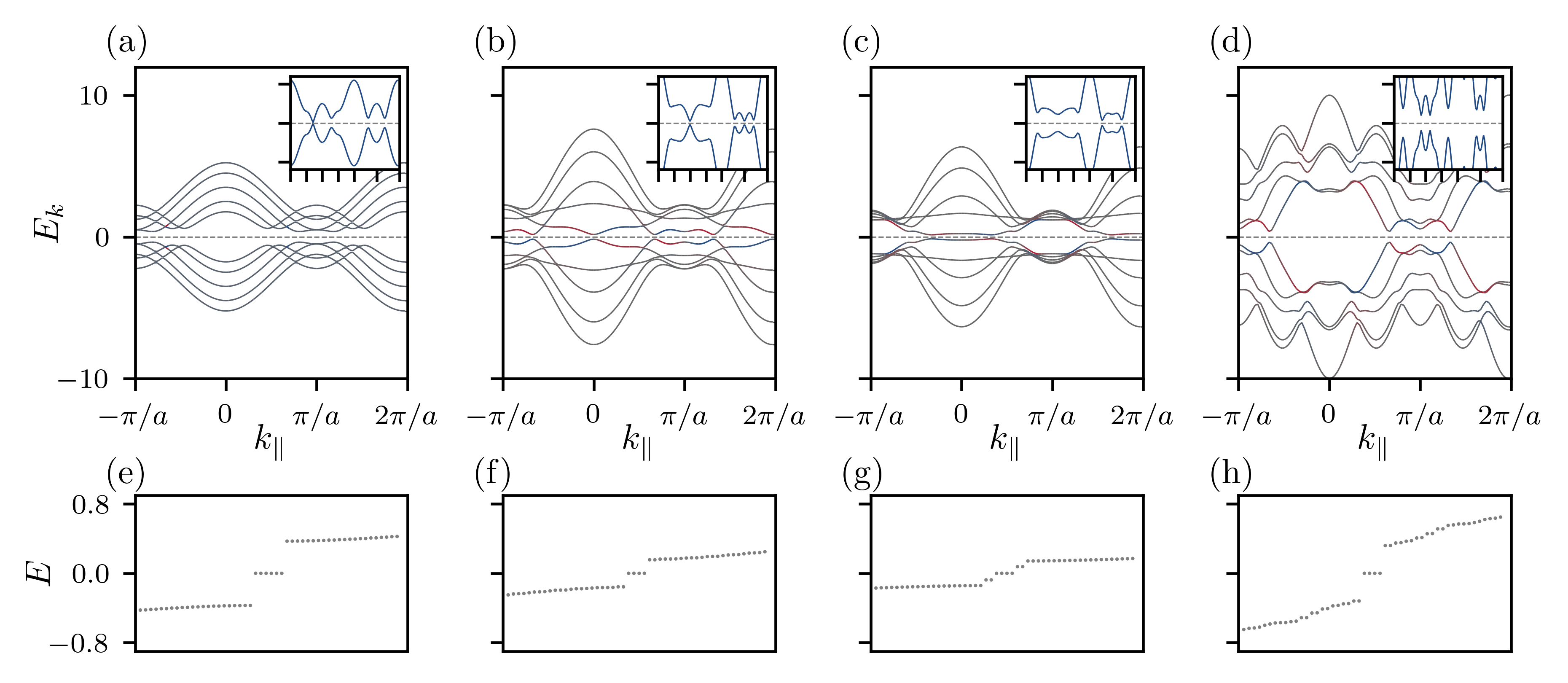}
        \caption{The energy bands of 5-chain-wide rectangular nanoribbons that illustrate the different methods to obtain higher winding number. (a) A non-extended model ($c_{n>1}=0$) with $\Delta_\perp=0$, resulting in effectively decoupled chains with renormalized masses, three of which are in the 1D topological phase, $C=0$ and $w=3$. $M=1.5t$ and $t_\parallel=t_\perp=\Delta_\parallel=1$ (b) A nanoribbon with parameters such that the 2D bulk has $C=2$, where both modes hybridize, resulting in $w=2$. $c_2=1$, $t_\parallel = t_\perp=1$, $\Delta_\parallel=\Delta_\perp=0.2t$, $t^{(2)}=0.4t$, $\Delta^{(2)}=0.6t$ and $M=t$. (c) A nanoribbon with parameters such that the 2D bulk has $C=1$, and the ribbon Hamiltonian has $w=2$. $c_2=1$, $t_\parallel = t_\perp=t$, $\Delta_\parallel=\Delta_\perp=0.2t$, $t^{(2)}=0.3t$, $\Delta^{(2)}=0.3t$ and $M=0.5t$. (d) A 2D trivial, but 1D topological realization of a nanoribbon, with $c_3=1$, $t_\parallel = t_\perp=\Delta_\parallel=\Delta_\perp=t$, $t^{(3)}=t$, $\Delta^{(3)}=0.6t$ and $M=0.5t$. Here $C=0$ but $w=2$.} 
        \label{fig:mechanisms}
    \end{figure*}

    \section{\label{sec:extended} Higher winding numbers \& extended models}
    When $\Delta_\perp = 0$, Eq.~\eqref{eq:PhaseBoundary} reduces to
    \[ M  = 2t_\perp \cos(q) \mp2t_\parallel,\]
    and the chains effectively decouple into independent Kitaev chains, with the phase boundaries given in terms of renormalized masses \cite{wakatsuki_majorana_2014}. 
    In this limit, each chain can host its own topological states, and the topological invariant is the sum of the topological invariants of the whole system.
    On the other hand, when $\Delta_\perp \neq 0$, the topological sectors couple and one has to calculate the invariant using Eq.~\eqref{eq:winding}. In turns out, that for the non-extended QWZ model with $\Delta_\perp \neq 0$, the winding number never exceeds one and the Pfaffian method captures all topological phases. However, inspired by the higher winding numbers achieved in extended SSH models, we will consider extended models which have either diagonal NNN ($c_2\neq0$) or NNNN $\Delta$ contributions ($c_3\neq0$), as depicted in Figs.~\ref{fig:5wideIntro}(a) and (d). 
    For the the nanoribbons, depicted in Figs.~\ref{fig:extendedHoppings}(b) and (e), indeed phases with higher winding number found. Note that the 4-chain-wide parallel sharp and the diagonal rectangular terminated nanoribbon do not admit a winding number. Figs.~\ref{fig:extendedHoppings}(c) and (f) corroborate that the 2D phase diagrams also exhibits higher Chern numbers, indicating regions with three topological edge modes.

    Upon the further inclusion of higher order hoppings, more diverse combinations of Chern numbers and winding numbers are possible. In Appendix \ref{app:ExtendedPhases}, Fig.~\ref{fig:extendedHoppings} is reproduced for $t^{(n)}=\Delta^{(n)}=0.3t$. 
    
    In Fig.~\ref{fig:mechanisms}, we present specific cases in which different combinations of Chern numbers and winding numbers occur. These are indicative of different mechanisms for driving dimensional crossovers leading to topological nanoribbons. Fig.~\ref{fig:mechanisms} illustrate four mechanisms in which nanoribbons acquire higher winding number.
    
    Firstly, in Fig.~\ref{fig:mechanisms}(a), the energy bands of a nanoribbon with no pairing between chains ($\Delta_\perp=0$) are shown. The Hamiltonian of this a system decouples into multiple chains with renormalized parameters, which independently go in and out of their topological phases. In general, the Hamiltonian can decompose in multiple Hamiltonians of nanoribbons of smaller width with renormalized parameters, which can be in a $w=1$ simultaneously. This is similar to stacking multiple topological insulators and adding their respective topological invariants, and is not a consequence of hybridization. 
    
    Additionally, Fig.~\ref{fig:mechanisms}(b) depicts a nanoribbon with parameters for which the 2D bulk has $C=2$ and consequently hosts 2 pairs of edge modes, as depicted in Fig.~\ref{fig:mechanisms}(f). In general, a nanoribbon can have $n$ 1D edge modes, which independently hybridize, resulting in a gapped system with $n$ pairs of topological end modes.
    
    Furthermore, there can be a single edge mode $C=1$ going through multiple avoided crossings, where each avoided crossing constitutes a topological gap and contributes to the winding number. This is illustrated in Fig.~\ref{fig:mechanisms}(c). Here, a single edge mode goes through two avoided crossings, each of which contributes one set of topological end modes.

    Lastly, Fig.~\ref{fig:mechanisms}(d) illustrates that even a system which is trivial in the 2D bulk, can host topological end states. Despite the lack of topological edge states in 2D, edge-localized solutions appear, which after hybridization do result in topologically protected end states.
    
    We want to emphasize that the first effect is not a dimensional crossover and is well-known in the literature. 
    However, the other effects are due to the hybridization of topological edge modes and arise in this dimensional transition. It is remarkable that a surprisingly simple model, nanoribbons described by the QWZ model, already exhibit a distinct set of mechanisms to achieve higher winding numbers. 
    However, because these larger number of topological end states are protected by the chiral symmetry, they only occur for specific terminations, often showing an even/odd effect in the width. Furthermore, their protection is weaker, because the protecting symmetry depends on the real-space geometry of the lattice. 
    
    \section{Conclusion \label{sec:Conclusions}}
    We have investigated the emergence of topological phases in 1D nanoribbons described by the QWZ model and its extensions with longer-range couplings. For the non-extended QWZ model, we provide analytic expressions for the gap-closing conditions and the corresponding topological phase boundaries, in line with the literature. These reveal how the hybridization of edge states produces a sequence of alternating trivial and topological gaps as the system parameters are varied. When the emergent chiral symmetry is present, the effective 1D system belongs to the symmetry class BDI, which admits an integer winding number; otherwise the classification reduces to the $\mathbb Z_2$-invariant associated with class D.

    We demonstrate that the topology of the hybridization gap is strongly influenced by the geometry of the nanoribbon. While PHS is inherited directly from the parent QWZ model, an additional chiral symmetry can emerge, strongly dependent on the ribbon termination. This symmetry combines orbital and spatial transformations and therefore has some crystalline character. Importantly, its existence is not universal: for most ribbon geometries it depends on the ribbon width, giving rise to an even-odd effect in the topological classification.
    
    We further explored extended QWZ models containing longer-range hopping and pairing terms. These extensions generate richer topological phase diagrams and allow the realization of phases with multiple topological end modes and higher winding numbers. We identified several mechanisms that can produce such phases, including the coexistence and hybridization of multiple edge modes inherited from higher-Chern-number bulk phases. Additionally, we also showed that there exists a parameter range in which the bulk host a single edge mode, but the nanoribbons have multiple edge modes. We argue that these $w>C$ phases occur due to multiple hybridization events at different crystal momentum.  These mechanisms illustrate that higher winding numbers do not need to originate from a single universal process but can emerge from qualitatively different forms of dimensional crossover.
    
    Taken together, our results show that the topology of confined CI and TSC systems is determined not only by the 2D bulk band structure but also by the geometry of the ribbon. Boundary termination, ribbon width, and finite-size hybridization can generate emergent crystalline symmetries and topological phases. We expect these findings to extend beyond the QWZ model, opening the door to understand and engineer topological phase by confining higher-dimensional topological materials.
    
	\begin{acknowledgments}
    L.E.~and C.M.S.~acknowledge the research program “Materials for the Quantum Age” (QuMat) for financial support. This program (registration number 024.005.006) is part of the Gravitation program financed by the Dutch Ministry of Education, Culture and Science (OCW).
	\end{acknowledgments}

    \appendix
    \section{Phase diagram with higher order hoppings couplings}\label{app:ExtendedPhases}
    \begin{figure*}
        \centering
        \includegraphics[]{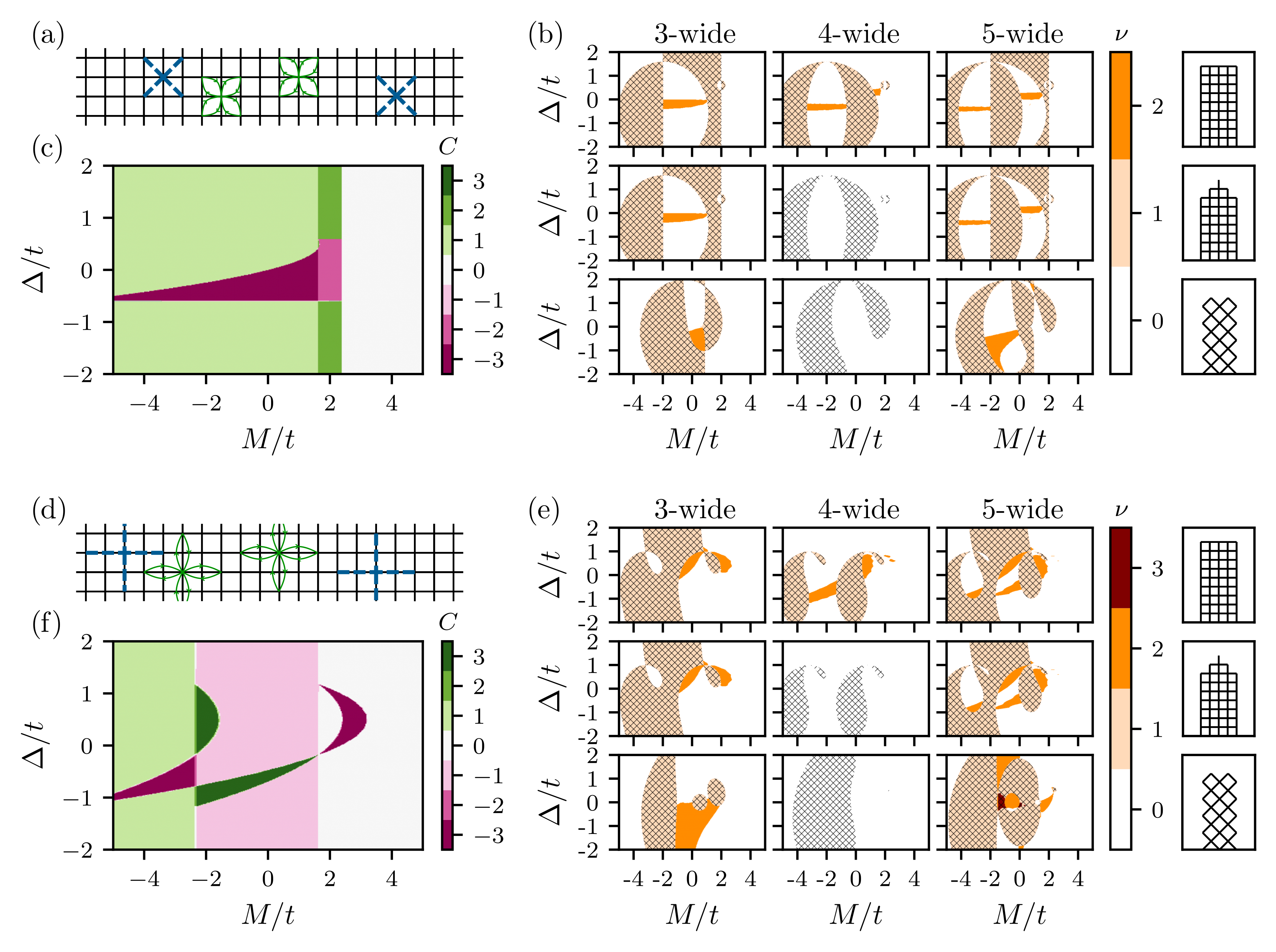}
        \caption{The topological phase space of extended QWZ nanoribbons including longer range hopping. ($t_\parallel=t_\perp=t,\Delta_\parallel=\Delta_\perp=\Delta$, $\Delta^{(n)} = t^{(n)} =0.2t$), with (a), (b) and (c) including NNN hopping $c_2 = 1$ and $c_{n\geq3}=0$ and (d), (e) and (f) including parallel NNNN hopping $c_3 =1$ and $c_{n\neq3}=0$. (a) indicates what extended couplings are considered.  (c) shows the phase space of the nanoribbons, for 3-, 4-, and 5-site-wide parallel rectangular, parallel sharp and diagonal rectangular nanoribbons. Here, the an non-zero Pfaffian invariant is indicated by hatched regions, and the winding number is captured by the colors. (c) the 2D phase diagram showing a non-zero Chern number when $|M|<4t$. (d), (e) and (f) are similar but with NNNN ($c_3 \neq 0 $) hopping instead of NNN hopping.}
        \label{fig:placeholder}
    \end{figure*}
    In Fig.~\ref{fig:placeholder}, we present the slices of the phase space of the 2D bulk and nanoribbons described by the extended QWZ model, taking into account both longer range SOC (CI) or pairing (TSC), and longer range hopping. It's structure is analogous to Fig.~\ref{fig:phase} of the main text. Again, Figs.~\ref{fig:placeholder}(a) and (d) indicate the extra terms on select sites. Figs.~\ref{fig:placeholder}(b) and (e), show the topological phases of nanoribbons of different termination as function of the Mass and the NN $\Delta$ coupling. Note that the 4-chain-wide parallel sharp and the diagonal rectangular terminated nanoribbon do not admit a winding number.
    Furthermore, the previous $M \rightarrow -M$ symmetry is lost. Figs.~\ref{fig:placeholder}(c) and (f), show the Chern number of the 2D bulk. Agian, the $M \rightarrow -M$ symmetry is lost. Remarkably, it has regions in which the winding number exceeds the Chern number, at the cost of a more complicated phase space. The mechanisms that result in $w>C$, depicted in Figs.~\ref{fig:mechanisms}(c) and (d), require this longer range kinetic term and are discussed in the main text.
	
    \bibliography{Master.bib} % Produces the bibliography via BibTeX.

@article{ryu_topological_2010,
	title = {Topological Insulators and Superconductors: Tenfold Way and Dimensional Hierarchy},
	volume = {12},
	issn = {1367-2630},
	shorttitle = {Topological Insulators and Superconductors},
	doi = {10.1088/1367-2630/12/6/065010},
	number = {6},
	journal = {New J. Phys.},
	author = {Ryu, Shinsei and Schnyder, Andreas P and Furusaki, Akira and Ludwig, Andreas W W},
	month = jun,
	year = {2010},
	pages = {065010},
}

@article{hasan_colloquium_2010,
	title = {Colloquium: Topological insulators},
	volume = {82},
	doi = {10.1103/RevModPhys.82.3045},
	number = {4},
	journal = {Rev. Mod. Phys.},
	author = {Hasan, M. Z. and Kane, C. L.},
	month = nov,
	year = {2010},
	pages = {3045--3067},
}

@article{rizzo_topological_2018,
	title = {Topological band engineering of graphene nanoribbons},
	volume = {560},
	copyright = {2018 Springer Nature Limited},
	issn = {1476-4687},
	doi = {10.1038/s41586-018-0376-8},
	number = {7717},
	journal = {Nature},
	author = {Rizzo, Daniel J. and Veber, Gregory and Cao, Ting and Bronner, Christopher and Chen, Ting and Zhao, Fangzhou and Rodriguez, Henry and Louie, Steven G. and Crommie, Michael F. and Fischer, Felix R.},
	month = aug,
	year = {2018},
	pages = {204--208},
}

@article{klaassen_realization_2025,
	title = {Realization of a one-dimensional topological insulator in ultrathin germanene nanoribbons},
	volume = {16},
	journal = {Nat Commun},
	author = {Klaassen, Dennis J. and Eek, Lumen and Rudenko, Alexander N. and van 't Westende, Esra D. and Castenmiller, Carolien and Zhang, Zhiguo and de Boeij, Paul L. and van Houselt, Arie and Ezawa, Motohiko and Zandvliet, Harold J. W. and Morais Smith, Cristiane and Bampoulis, Pantelis},
	year = {2025},
	pages = {2059},
}

@article{eek_electric-field_2025,
	title = {Electric-Field Control of Zero-Dimensional Topological States in Ultranarrow Germanene Nanoribbons},
	volume = {135},
	doi = {10.1103/jx2x-fb5b},
	number = {20},
	journal = {Phys. Rev. Lett.},
	publisher = {American Physical Society},
	author = {Eek, Lumen and van 't Westende, Esra D. and Klaassen, Dennis J. and Zandvliet, Harold J. W. and Bampoulis, Pantelis and Smith, Cristiane Morais},
	month = nov,
	year = {2025},
	pages = {206601},
}

@article{chiu_classification_2016,
	title = {Classification of topological quantum matter with symmetries},
	volume = {88},
	doi = {10.1103/RevModPhys.88.035005},
	number = {3},
	journal = {Rev. Mod. Phys.},
	publisher = {American Physical Society},
	author = {Chiu, Ching-Kai and Teo, Jeffrey C. Y. and Schnyder, Andreas P. and Ryu, Shinsei},
	month = aug,
	year = {2016},
	pages = {035005},
}

@article{molinari_determinants_2008,
	title = {Determinants of block tridiagonal matrices},
	volume = {429},
	copyright = {https://www.elsevier.com/tdm/userlicense/1.0/},
	issn = {00243795},
	doi = {10.1016/j.laa.2008.06.015},
	number = {8-9},
	journal = {Linear Algebra and its Applications},
	author = {Molinari, Luca Guido},
	month = oct,
	year = {2008},
	pages = {2221--2226},
}

@article{wakatsuki_majorana_2014,
	title = {Majorana fermions and multiple topological phase transition in {K}itaev ladder topological superconductors},
	volume = {89},
	copyright = {http://link.aps.org/licenses/aps-default-license},
	issn = {1098-0121, 1550-235X},
	doi = {10.1103/PhysRevB.89.174514},
	number = {17},
	journal = {Phys. Rev. B},
	author = {Wakatsuki, Ryohei and Ezawa, Motohiko and Nagaosa, Naoto},
	month = may,
	year = {2014},
	pages = {174514},
}

@misc{osseweijer_topology_2026,
	title = {Topology of honeycomb nanoribbons revisited},
	copyright = {arXiv.org perpetual, non-exclusive license},
	doi = {10.48550/ARXIV.2603.25497},
	publisher = {arXiv},
	author = {Osseweijer, Zebedeus F. and Eek, Lumen and Zandvliet, Harold J. W. and Bampoulis, Pantelis and Smith, Cristiane Morais},
	year = {2026},
}

@article{maisel_liceran_topology_2024,
	title = {Topology of {Bi$_2$Se$_3$} nanosheets},
	volume = {109},
	issn = {2469-9950, 2469-9969},
	doi = {10.1103/PhysRevB.109.195407},
	number = {19},
	journal = {Phys. Rev. B},
	author = {Maisel Licerán, L. and Koerhuis, S. J. H. and Vanmaekelbergh, D. and Stoof, H. T. C.},
	month = may,
	year = {2024},
	pages = {195407},
}

@article{liu_oscillatory_2010,
	title = {Oscillatory crossover from two-dimensional to three-dimensional topological insulators},
	volume = {81},
	copyright = {http://link.aps.org/licenses/aps-default-license},
	issn = {1098-0121, 1550-235X},
	doi = {10.1103/PhysRevB.81.041307},
	number = {4},
	journal = {Phys. Rev. B},
	author = {Liu, Chao-Xing and Zhang, HaiJun and Yan, Binghai and Qi, Xiao-Liang and Frauenheim, Thomas and Dai, Xi and Fang, Zhong and Zhang, Shou-Cheng},
	month = jan,
	year = {2010},
	pages = {041307},
}

@article{singh_topological_2013,
	title = {Topological phase transition and two-dimensional topological insulators in {Ge}-based thin films},
	volume = {88},
	copyright = {http://link.aps.org/licenses/aps-default-license},
	issn = {1098-0121, 1550-235X},
	doi = {10.1103/PhysRevB.88.195147},
	number = {19},
	journal = {Phys. Rev. B},
	author = {Singh, Bahadur and Lin, Hsin and Prasad, R. and Bansil, A.},
	month = nov,
	year = {2013},
	pages = {195147},
}

@article{qi_topological_2006,
	title = {Topological quantization of the spin {H}all effect in two-dimensional paramagnetic semiconductors},
	volume = {74},
	copyright = {http://link.aps.org/licenses/aps-default-license},
	issn = {1098-0121, 1550-235X},
	doi = {10.1103/PhysRevB.74.085308},
	number = {8},
	journal = {Phys. Rev. B},
	author = {Qi, Xiao-Liang and Wu, Yong-Shi and Zhang, Shou-Cheng},
	month = aug,
	year = {2006},
	pages = {085308},
}

@article{read_paired_2000,
	title = {Paired states of fermions in two dimensions with breaking of parity and time-reversal symmetries and the fractional quantum {H}all effect},
	volume = {61},
	copyright = {http://link.aps.org/licenses/aps-default-license},
	issn = {0163-1829, 1095-3795},
	doi = {10.1103/PhysRevB.61.10267},
	number = {15},
	journal = {Phys. Rev. B},
	author = {Read, N. and Green, Dmitry},
	month = apr,
	year = {2000},
	pages = {10267--10297},
}

@misc{balling-anso_transition_2026,
	title = {Transition between one- and two-dimensional topology in a {C}hern insulator of finite width},
	copyright = {arXiv.org perpetual, non-exclusive license},
	doi = {10.48550/ARXIV.2602.16411},
	publisher = {arXiv},
	author = {Balling-Ansø, Frode and Pal, Adipta and Cook, Ashley M. and Nielsen, Anne E. B.},
	year = {2026},
}

@article{tewari_topological_2012,
	title = {Topological minigap in quasi-one-dimensional spin-orbit-coupled semiconductor {M}ajorana wires},
	volume = {86},
	copyright = {http://link.aps.org/licenses/aps-default-license},
	issn = {1098-0121, 1550-235X},
	doi = {10.1103/PhysRevB.86.024504},
	number = {2},
	journal = {Phys. Rev. B},
	author = {Tewari, Sumanta and Stanescu, T. D. and Sau, Jay D. and Das Sarma, S.},
	month = jul,
	year = {2012},
	pages = {024504},
}

@article{cook_finite-size_2023,
	title = {Finite-size topology},
	volume = {108},
	issn = {2469-9950, 2469-9969},
	doi = {10.1103/PhysRevB.108.045144},
	number = {4},
	journal = {Phys. Rev. B},
	author = {Cook, Ashley M. and Nielsen, Anne E. B.},
	month = jul,
	year = {2023},
	pages = {045144},
}

@article{potter_multichannel_2010,
	title = {Multichannel Generalization of {K}itaev’s {M}ajorana End States and a Practical Route to Realize Them in Thin Films},
	volume = {105},
	copyright = {http://link.aps.org/licenses/aps-default-license},
	issn = {0031-9007, 1079-7114},
	doi = {10.1103/PhysRevLett.105.227003},
	number = {22},
	journal = {Phys. Rev. Lett.},
	author = {Potter, Andrew C. and Lee, Patrick A.},
	month = nov,
	year = {2010},
	pages = {227003},
}

@article{wang_topological_2018,
	title = {Topological phase diagrams and {M}ajorana zero modes of the {K}itaev ladder and tube},
	volume = {27},
	copyright = {http://iopscience.iop.org/info/page/text-and-data-mining},
	issn = {1674-1056},
	doi = {10.1088/1674-1056/27/4/047401},
	number = {4},
	journal = {Chinese Phys. B},
	author = {Wang, Yiming and Li, Zhidan and Han, Qiang},
	month = apr,
	year = {2018},
	pages = {047401},
}

@article{traverso_emerging_2024,
	title = {Emerging topological bound states in {Haldane} model zigzag nanoribbons},
	volume = {9},
	doi = {10.1038/s41535-023-00615-1},
	number = {1},
	urldate = {2026-06-01},
	journal = {npj Quantum Mater.},
	author = {Traverso, Simone and Sassetti, Maura and Traverso Ziani, Niccolò},
	month = jan,
	year = {2024},
	pages = {9},
}

@article{zhao_topological_2021,
	title = {Topological phases in Graphene Nanoribbons Tuned by Electric Fields},
	volume = {127},
	issn = {0031-9007, 1079-7114},
	url = {https://link.aps.org/doi/10.1103/PhysRevLett.127.166401},
	doi = {10.1103/PhysRevLett.127.166401},
	number = {16},
	urldate = {2026-06-01},
	journal = {Phys. Rev. Lett.},
	author = {Zhao, Fangzhou and Cao, Ting and Louie, Steven G.},
	month = oct,
	year = {2021},
	pages = {166401},
}

@article{hsieh_observation_2009,
	title = {Observation of Time-Reversal-Protected Single-{Dirac}-Cone Topological-Insulator States in {Bi$_2$Te$_3$}and {Sb$_2$Te$_3$}},
	volume = {103},
	doi = {10.1103/PhysRevLett.103.146401},
	number = {14},
	urldate = {2026-06-01},
	journal = {Phys. Rev. Lett.},
	author = {Hsieh, D. and Xia, Y. and Qian, D. and Wray, L. and Meier, F. and Dil, J. H. and Osterwalder, J. and Patthey, L. and Fedorov, A. V. and Lin, H. and Bansil, A. and Grauer, D. and Hor, Y. S. and Cava, R. J. and Hasan, M. Z.},
	month = sep,
	year = {2009},
	pages = {146401},
}

@article{zhang_edge_2015,
	title = {Edge states and integer quantum {Hall} effect in topological insulator thin films},
	volume = {5},
	issn = {2045-2322},
	doi = {10.1038/srep13277},
	number = {1},
	urldate = {2026-06-01},
	journal = {Sci Rep},
	author = {Zhang, Song-Bo and Lu, Hai-Zhou and Shen, Shun-Qing},
	month = aug,
	year = {2015},
	pages = {13277},
}

@article{linder_anomalous_2009,
	title = {Anomalous finite size effects on surface states in the topological insulator {Bi$_2$Se$_3$}},
	volume = {80},
	url = {https://link.aps.org/doi/10.1103/PhysRevB.80.205401},
	doi = {10.1103/PhysRevB.80.205401},
	number = {20},
	urldate = {2026-06-01},
	journal = {Phys. Rev. B},
	author = {Linder, Jacob and Yokoyama, Takehito and Sudbø, Asle},
	month = nov,
	year = {2009},
	pages = {205401},
}

@article{shan_effective_2010,
	title = {Effective continuous model for surface states and thin films of three-dimensional topological insulators},
	volume = {12},
	issn = {1367-2630},
	url = {https://iopscience.iop.org/article/10.1088/1367-2630/12/4/043048},
	doi = {10.1088/1367-2630/12/4/043048},
	number = {4},
	urldate = {2026-06-01},
	journal = {New J. Phys.},
	author = {Shan, Wen-Yu and Lu, Hai-Zhou and Shen, Shun-Qing},
	month = apr,
	year = {2010},
	pages = {043048},
}

@article{kitaev_unpaired_2001,
	title = {Unpaired {Majorana} fermions in quantum wires},
	volume = {44},
	issn = {1468-4780},
	doi = {10.1070/1063-7869/44/10S/S29},
	number = {10S},
	urldate = {2026-06-01},
	journal = {Phys.-Usp.},
	author = {Kitaev, A Yu},
	month = oct,
	year = {2001},
	pages = {131--136},
}

@article{zak_berrys_1989,
	title = {Berry’s phase for energy bands in solids},
	volume = {62},
	copyright = {http://link.aps.org/licenses/aps-default-license},
	issn = {0031-9007},
	url = {https://link.aps.org/doi/10.1103/PhysRevLett.62.2747},
	doi = {10.1103/PhysRevLett.62.2747},
	number = {23},
	urldate = {2025-09-24},
	journal = {Phys. Rev. Lett.},
	author = {Zak, J.},
	month = jun,
	year = {1989},
	pages = {2747--2750},
}

\end{document}